# Approximate Solutions, Thermal Properties and Superstatistics Solutions to Schrödinger Equation

I. B. Okon[1*], C. A. Onate[2], E. Omugbe[3], U.S. Okorie[4], A. D. Antia[1], M. C. Onyeaju[5], Chen Wen-Li[6] and J. P. Araujo[7]

[1]Theoretical Physics Group, Department of Physics, University of Uyo, Nigeria.
[2]Department of Physical Sciences, Landmark University, Omu-Aran, Nigeria.
[3]Department of Physics, Federal University of Petroleum Resources, Effurun, Nigeria.
[4]Department of Physics, Akwa Ibom State University, Ikot Akpaden, Uyo, Nigeria.
[5]Theoretical Physics Group, Department of Physics, University of Port Harcourt, Nigeria.
[6]School of Intelligent Science and Information Engineering, Xi'an Peihua University, Xi'an, China.
[7]Department of Mathematics, Federal Institute of the Southeast of Minas Gerais, Juiz de Fora, Brazil.

The corresponding author email: *ituenokon@uniuyo.edu.ng

**Abstract**
In this work, we apply the parametric Nikiforov-Uvarov method to obtain eigen solutions and total normalized wave function of Schrödinger equation express in terms of Jacobi polynomial using Coulomb plus Screened Exponential Hyperbolic potential (CPSEHP), where we obtained the probability density plots for the proposed potential for various orbital angular quantum number, as well as some special cases (Hellmann and Yukawa potential). The proposed potential is best suitable for smaller values of the screening parameter $\alpha$. The resulting energy eigen equation is presented in a close form and extended to study thermal properties and superstatistics express in terms of partition function $(Z)$ and other thermodynamic properties such as; vibrational mean energy $(U)$, vibrational specific heat capacity $(C)$, vibrational entropy $(S)$ and vibrational free energy $(F)$. Using the resulting energy equation and with the help of Matlab software, the numerical bound state solutions were obtained for various values of the screening parameter ($\alpha$) as well as different expectation values via Hellmann-Feynman Theorem (HFT). The trend of the partition function and other thermodynamic properties obtained for both thermal properties and superstatistics were in excellent agreement with the existing literatures. Due to the analytical mathematical complexities, the superstatistics and thermal properties were evaluated using Mathematica 10.0 version software. The proposed potential model reduces to Hellmann potential, Yukawa potential, Screened Hyperbolic potential and Coulomb potential as special cases.

## 1. Introduction

The approximate analytical solutions of one-dimensional radial Schrödinger wave equation with a multiple potential function has been studied using a suitable approximation scheme to the centrifugal term within the frame work of parametric Nikiforov-Uvarov method [1]. The solutions to the wave equations in quantum mechanics and applied physics play crucial role in understanding the importance of physical systems [2]. The two most important parts in studying Schrödinger wave equations are the total wave function and energy eigenvalues [3]. The analytic solutions of wave equations for some physical potentials are possible for $l = 0$. For $l \neq 0$, special approximation scheme like the Greene- Aldrich and Pekeris approximations are employ to deal with the centrifugal barrier in order to obtain an approximate bound state solutions [4-6]. The Greene-Aldrich approximation scheme is mostly applicable for short range potentials [7]. Eigen solutions for both relativistic and nonrelativistic wave equations have been studied with different methods which include: Exact quantisation, WKB, Nikiforov-Uvarov method (NU), Laplace transform technique, asymptotic iteration method, proper quantisation, supersymmetric quantum mechanics approach, vibrational approach, formula method, factorisation method, and the Shifted 1/N-expansion method [8-13]. Bound state solutions obtained from Schrodinger wave equation has practical applications in investigating tunnelling rate of quantum mechanical systems [14] and mass spectra of quarkonia systems [15-19]. Among other goals achieve in this research article is to apply the Hellmann-Feynman Theorem (HFT) to eigen equation of the Schrödinger wave equations to obtain expectation values of $<r^{-1}>_{nl}, <r^{-2}>_{nl}$ , $<T>_{nl}$ and $<p^2>_{nl}$ analytically. The Hellmann-Feynman Theorem gives an insight about chemical bonding and other forces existing among atoms of molecules [20-25]. To engage HFT in calculating the expectation values, one needs to promote the fixed parameter which appears in the Hamiltonian to be continuous variable in order to ease the mathematical purpose of taking the derivative [26]. Similarly, application of Hellmann-Feynman Theorem provides a less mathematical approach of obtaining expectation values of a quantum mechanical systems [27-28]. Some of the potential models considered within the framework of relativistic and nonrelativistic wave equations are: Hulthen-Yukawa Inversely quadratic potential [29], noncentral Inversely quadratic potential [30], Modified Hylleraas potential [31], Yukawa, Hulthen, Eckart, Deng-Fan, Pseudoharmonic , Kratzer, Woods-saxon, double ring shape , Coulomb, Tietz –wei , Tietz-Hua , Deng-Fan, Manning-Rosen , trigonometric Rosen-Morse, hyperbolic scalar and vector potential and exponential type potentials among others [32-50]. Coulomb, hyperbolic and screened exponential type potentials have been of interest to researchers in recent times because of their enormous applications in both chemical and physical sciences. In view of this, Parmar [51] studied ultra-generalized exponential hyperbolic potential where he obtained energy eigenvalues, un-normalised wave function and the partition function. This potential reduces to Yukawa potential, Screened cosine kratzer potential, Manning-Rosen potential, Hulthen plus Inversely quadratic exponential Mie-type potential and many others. Okon et al. [52], in their studies, obtained eigen solutions to Schrödinger equation with trigonometric Inversely quadratic plus Coulombic hyperbolic potential where they obtained energy eigen equation and normalised wave function using Nikiforov-Uvarov method. Onate [53] examined bound state solutions of the Schrödinger equation with second Poschl-Teller-like potential where he obtained vibrational partition function, mean energy, vibrational specific heat capacity and mean free energy. In that work, the Poschl-Teller like potential was expressed in (hyperbolic form). The practical application of energy eigen equation of Schrödinger equation in investigating the partition function, thermodynamic properties and superstatistics arouses the interest of many researchers. Recently, Okon et al. [54], obtained the thermodynamic properties and bound state solutions of the Schrödinger equation using Mobius square plus screened  Kratzer potential for two diatomic systems (Carbon(II) oxide and Scandium Flouride) within the framework of Nikiforov-Uvarov method. Their results were in agreement to semi-classical WKB among others. They presented energy eigen equation in a close form in order to obtain partition function and other thermodynamic properties. Omugbe et al. [55], recently studied the unified treatment of the non-relativistic bound state solutions, thermodynamic properties and expectation values of exponential-type potentials where they obtained the thermodynamic properties within the framework of semi-classical WKB approach. The authors studied the special cases of the potential as Eckart, Manning-Rosen and Hulthén potentials. Besides, Oyewumi et al. [56], studied the thermodynamic properties and the approximate solutions of the Schrodinger equation with shifted Deng-Fan potential model within the framework of asymptotic Iteration method where they apply Perkeris-type approximation to centrifugal term to obtain rotational-vibrational energy eigenvalues for selected diatomic systems A lot of research have been carried out by A.N. Ikot, U. S Okorie and co-authors. These can be seen in Refs. [57-60]. Also, Boumali and Hassanabadi [61] studied thermal properties of a two dimensional Dirac oscillator under an external magnetic field where they obtained relativistic spin-1\2 fermions subject to Dirac oscillator coupling and a constant magnetic field in both commutative and non-commutative space.

In this work, we propose a novel potential called Coulomb plus screened Hyperbolic potential to study bound state solutions, expectation values, superstatistics and thermal properties within the framework of parametric Nikiforov-Uvarov method. This article is divided into 9 sections. The introduction is given in section 1. The parametric Nikiforov-Uvarov method is presented in section 2. The solutions of the radial Schrodinger wave equation is presented in section 3. The application of Hellmann–Feynman Theorem to obtain expectation values is presented in section 4. The thermodynamic properties and superstatistics formulations are presented in sections 5 and 6 respectively. Numerical results and discussion are presented sections 7 and 8 respectively, and the article is concluded in section 9

The propose Coulomb plus screened hyperbolic exponential potential (CPSHEP) is given as

$$V(r) = -\frac{v_1}{r} + \left(\frac{B}{r} - \frac{v_2 \cosh\alpha}{r^2}\right)e^{-\alpha r}, \quad (1)$$

where $v_1$ and $v_2$ are the potential depth, $B$ is a real constant parameter, $\alpha$ is the adjustable screening parameter. The Pekeries –like approximation to the centrifugal term is given as

$$\frac{1}{r^2} = \frac{\alpha^2}{\left(1-e^{-\alpha r}\right)^2} \Rightarrow \frac{1}{r} = \frac{\alpha}{\left(1-e^{-\alpha r}\right)}. \quad (2)$$

The graph of Pekeris approximation to centrifugal term is given below

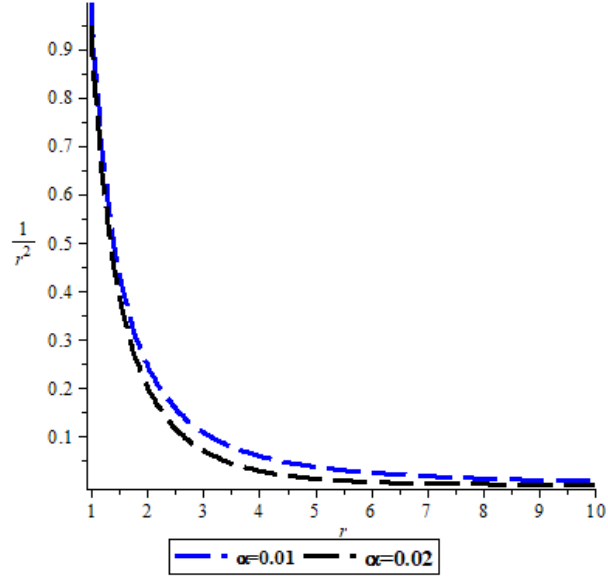

Figure 1: The graph of Pekeries approximation for various values of $\alpha$

## 2. Parametric Nikiforov-Uvarov (NU) Method

The NU method is based on reducing second order linear differential equation to a generalized equation of hyper-geometric type and provides exact solutions in terms of special orthogonal functions like Jacobi and Laguerre as well as corresponding energy eigenvalues [62-69]. The reference equation for parametric NU method according to Tezcan and Sever [70] is given as

$$\Psi''(s) + \frac{c_1 - c_2 s}{s(1 - c_3 s)}\Psi'(s) + \frac{1}{s^2(1 - c_3 s)^2}\left[-\Omega_1 s^2 + \Omega_2 s - \Omega_3\right]\Psi(s) = 0. \tag{3}$$

The condition for energy equation is given as [70]

$$c_2 n - (2n+1)c_5 + (2n+1)\left(\sqrt{c_9} + c_3\sqrt{c_8}\right) + n(n-1)c_3 + c_7 + 2c_3 c_8 + 2\sqrt{c_8 c_9} = 0. \tag{4}$$

The total wave function is given as

$$\psi_{nl}(s) = N_{nl} s^{c_{12}} (1 - c_3 s)^{-c_{12} - \frac{c_{13}}{c_3}} P_n^{\left(c_{10}-1, \frac{c_{11}}{c_3} - c_{10} - 1\right)}(1 - 2c_3 s). \tag{5}$$

The parametric constants are obtained as follows

$$c_1 = c_2 = c_3 = 1, c_4 = \frac{1}{2}(1 - c_1), \quad c_5 = \frac{1}{2}(c_2 - 2c_3), \quad c_6 = c_5^2 + \Omega_1$$

$$c_7 = 2c_4 c_5 - \Omega_2, \quad c_8 = c_4^2 + \Omega_3, \quad c_9 = c_3 c_7 + c_3^2 c_8 + c_6, \quad c_{10} = c_1 + 2c_4 + 2\sqrt{c_8}. \tag{6}$$

$$c_{11} = c_2 - 2c_5 + 2(\sqrt{c_9} + c_3\sqrt{c_8}), \quad c_{12} = c_4 + \sqrt{c_8}, \quad c_{13} = c_5 - \left(\sqrt{c_9} + c_3\sqrt{c_8}\right)$$

## 3. The Radial Solution of Schrodinger Wave Equation.

The radial Schrodinger wave equation with the centrifugal term is given as

$$\frac{d^2R(r)}{dr^2} + \frac{2\mu}{\hbar^2}\left[E - V(r) - \frac{\hbar^2 l(l+1)}{2\mu r^2}\right]R(r) = 0. \tag{7}$$

Equation (7) can only be solved analytically to obtain exact solution if the angular orbital quantum number $l = 0$. However, for $l > 0$ equation (7) can only be solve by using the approximations in (2) to the centrifugal term. Substituting equation (1) into (7) gives

$$\frac{d^2R(r)}{dr^2} + \frac{2\mu}{\hbar^2}\left[E_{nl} + \frac{v_1}{r} - \frac{Be^{-\alpha r}}{r} + \frac{v_2 e^{-\alpha r}\cosh\alpha}{r^2} - \frac{\hbar^2 l(l+1)}{2\mu r^2}\right]R(r) = 0. \tag{8}$$

By substituting equation (2) into (8) gives the following equation.

$$\frac{d^2R(r)}{dr^2} + \frac{2\mu}{\hbar^2}\left[E_{nl} + \frac{v_1\alpha}{(1-e^{-\alpha r})} - \frac{B\alpha e^{-\alpha r}}{(1-e^{-\alpha r})} + \frac{v_2\alpha^2 e^{-\alpha r}\cosh\alpha}{(1-e^{-\alpha r})^2} - \frac{\hbar^2\alpha^2 l(l+1)}{2\mu(1-e^{-\alpha r})^2}\right]R(r) = 0. \tag{9}$$

By defining $s = e^{-\alpha r}$, and with some simple algebraic simplification, equation (9) can be presented in the form

$$\frac{d^2R(s)}{ds^2} + \frac{(1-s)}{s(1-s)}\frac{dR}{ds} + \frac{1}{s^2(1-s)^2}\left\{\begin{array}{l}-(\varepsilon^2-\chi_1)s^2 \\ +(2\varepsilon^2-\delta^2-\chi_1+\chi_2)s-(\varepsilon^2-\delta^2+l(l+1))\end{array}\right\}R(s) = 0, \tag{10}$$

where

$$\varepsilon^2 = -\frac{2\mu E_{nl}}{\hbar^2\alpha^2}, \quad \delta^2 = \frac{2\mu v_1}{\hbar^2\alpha}, \quad \chi_1 = \frac{2\mu B}{\hbar^2\alpha}, \quad \chi_2 = \frac{2\mu v_2\cosh\alpha}{\hbar^2}. \tag{11}$$

Comparing equation (10) to (3), the following polynomials were obtain

$$\Omega_1 = (\varepsilon^2 - \chi_1), \quad \Omega_2 = (2\varepsilon^2 - \delta^2 - \chi_1 + \chi_2), \quad \Omega_3 = (\varepsilon^2 - \delta^2 + l(l+1)). \tag{12}$$

Using equation (6), other parametric constants are obtained as follows

$$c_1 = c_2 = c_3 = 1; c_4 = 0, c_5 = -\frac{1}{2}, \quad c_6 = \frac{1}{4} + \varepsilon^2 - \chi_1, c_7 = -2\varepsilon^2 + \delta^2 + \chi_1 - \chi_2$$

$$c_8 = \varepsilon^2 - \delta^2 + l(l+1), \quad c_9 = \frac{1}{4} - \chi_2 + l(l+1), \quad c_{10} = 1 + 2\sqrt{\varepsilon^2 - \delta^2 + l(l+1)}$$

$$c_{11} = 2 + \sqrt{1 - 4\chi_2 + 4l(l+1)} + 2\sqrt{\varepsilon^2 - \delta^2 + l(l+1)}, \quad c_{12} = \sqrt{\varepsilon^2 - \delta^2 + l(l+1)}$$

$$c_{13} = -\frac{1}{2} - \left[\frac{1}{2}\sqrt{1 - 4\chi_2 + 4l(l+1)} + \sqrt{\varepsilon^2 - \delta^2 + l(l+1)}\right]$$

(13)

Using equation (4), equation (12) and equation (13) with much algebraic simplification, the energy eigen equation for the proposed potential is given as

$$E_{nl} = \frac{\hbar^2 \alpha^2 l(l+1)}{2\mu} - v_1 \alpha + \frac{\hbar^2 \alpha^2}{2\mu} \left\{ \frac{\left(n^2 + n + \frac{1}{2}\right) + \left(n + \frac{1}{2}\right)\sqrt{1 + 4l(l+1) - \frac{8v_2 \mu \cosh\alpha}{\hbar^2}} + \frac{2\mu B}{\hbar^2 \alpha} - \frac{2\mu v_1}{\hbar^2 \alpha}}{(2n+1) + \sqrt{1 + 4l(l+1) - \frac{8v_2 \mu \cosh\alpha}{\hbar^2}}} \right\}^2$$

(14)

Using equation (5), the total un-normalised wave function is given as

$$\Psi_{nl}(s) = N_{nl} s^{\sqrt{(\varepsilon^2 - \delta^2 + l(l+1))}} (1-s)^{-\frac{1}{2} - 2\sqrt{(\varepsilon^2 - \delta^2 + l(l+1))} - \frac{1}{2}\sqrt{1 + 4l(l+1) - 4\chi_2}} P_n^{\left[\left(2\sqrt{(\varepsilon^2 - \delta^2 + l(l+1))}\right), \left(\sqrt{1 + 4l(l+1) - 4\chi_2}\right)\right]} (1 - 2s) \quad (15)$$

To obtain the normalization constant of equation (15), we employ the normalization condition

$$\int_0^\infty |R_{nl}(r)|^2 dr = 1 \Rightarrow \int_0^\infty \left[N_{nl} s^\beta (1-s)^\eta P_n^{(2\beta, 2\eta - 1)}(1 - 2s)\right]^2 ds = 1 \quad (16)$$

Where

$$\beta = \sqrt{\left(\varepsilon^2 - \delta^2 + l(l+1)\right)}, \quad (17)$$

$$\eta = \sqrt{1 + 4l(l+1) - 4\chi_2}. \quad (18)$$

The wave function is assumed to be in bound at $r \in (0, \infty)$ and $s = e^{-\alpha r} \in (1, 0)$

Equation (15) reduces to

$$-\frac{N_{nl}^2}{\alpha}\int_1^0 s^{2\beta}(1-s)^{2\eta}\left[P_n^{(2\beta,2\eta-1)}(1-2s)\right]^2 \frac{ds}{s} = 1. \qquad (19)$$

Let $z = (1-2s)$ such that the boundary of integration of equation (19) changes from $s \in (1, 0)$ to $z \in (-1, 1)$. Then equation (19) reduces to

$$\frac{N_{nl}^2}{2\alpha}\int_{-1}^1 \left(\frac{1-z}{2}\right)^{2\beta-1}\left(\frac{1+z}{2}\right)^{2\eta}\left[P_n^{(2\beta,2\eta-1)}(z)\right]^2 dz = 1. \qquad (20)$$

Using the standard integral

$$\int_{-1}^1 \left(\frac{1-w}{2}\right)^x \left(\frac{1+w}{2}\right)^y \left[P_n^{(x,y-1)}(w)\right]^2 dw = \frac{2^{x+y+1}\Gamma(x+n+1)\Gamma(y+n+1)}{n!\,\Gamma(x+y+n+1)\Gamma(x+y+2n+1)}. \qquad (21)$$

Let

$z = w$, $x = 2\beta - 1$, $y = 2\eta$. Then using equation (20) the normalization constant can be obtained as

$$N_{nl} = \sqrt{\frac{2\alpha(n!)\Gamma(2\beta+2\eta+n)\Gamma(2\beta+2\eta+2n)}{2^{(2\beta+2\eta)}\Gamma(2\beta+n)\Gamma(2\eta+n+1)}}. \qquad (22)$$

Hence the total normalized wave function is given as

$$R_{n,l}(s) = \sqrt{\frac{2\alpha(n!)\Gamma(2\beta+2\eta+n)\Gamma(2\beta+2\eta+2n)}{2^{(2\beta+2\eta)}\Gamma(2\beta+n)\Gamma(2\eta+n+1)}}\, s^\beta (1-s)^\eta P_n^{(2\beta,2\eta-1)}(1-2s). \qquad (23)$$

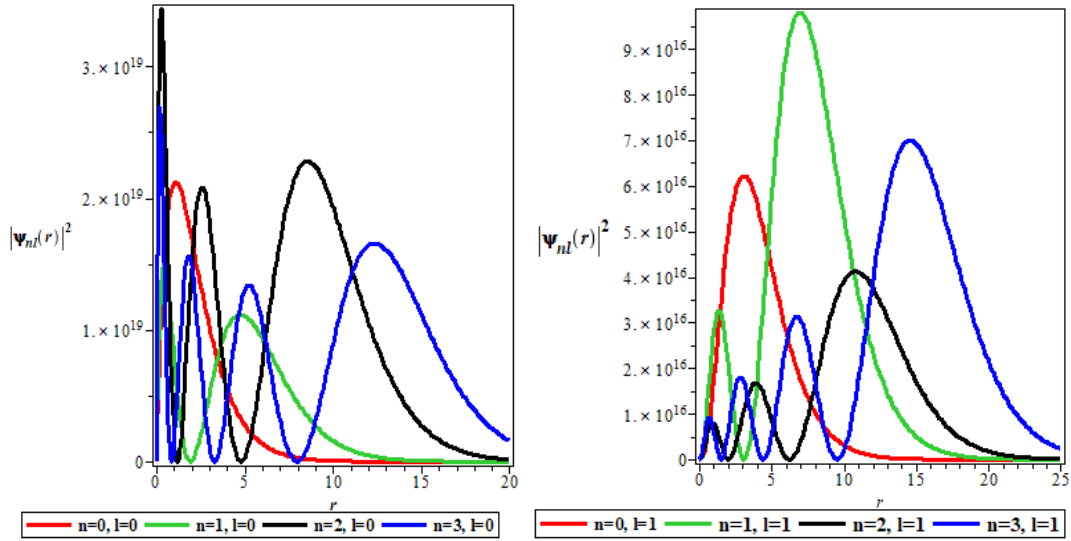

Figure 2: Variation of the probability density against the internuclear separation at various at various quantum states

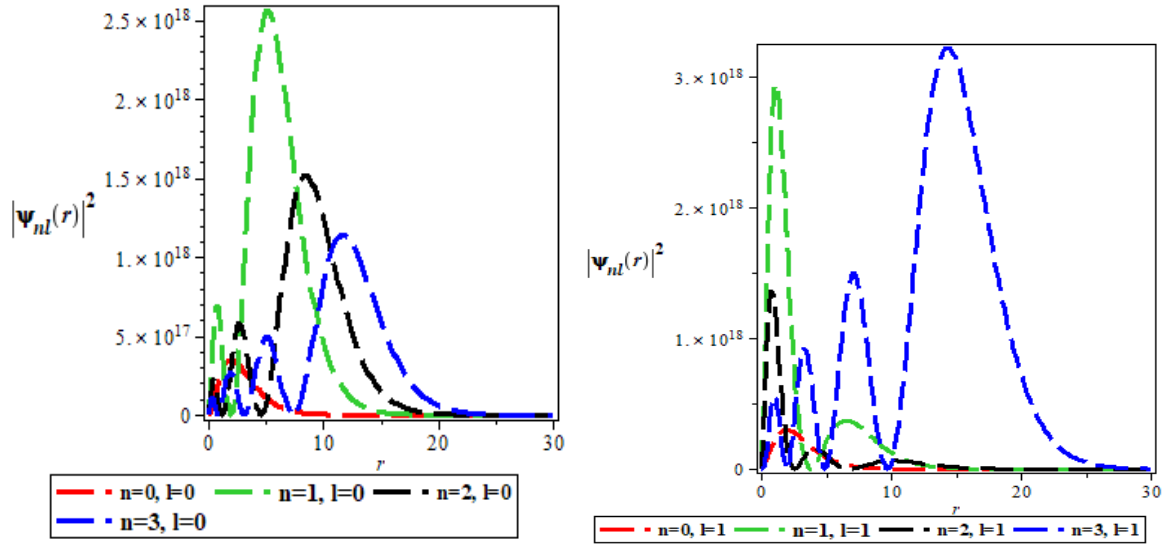

Figure 3: Variation of the probability density against the internuclear separation at various quantum state for $l=0$ and $l=1$ respectively for Hellmann potential.

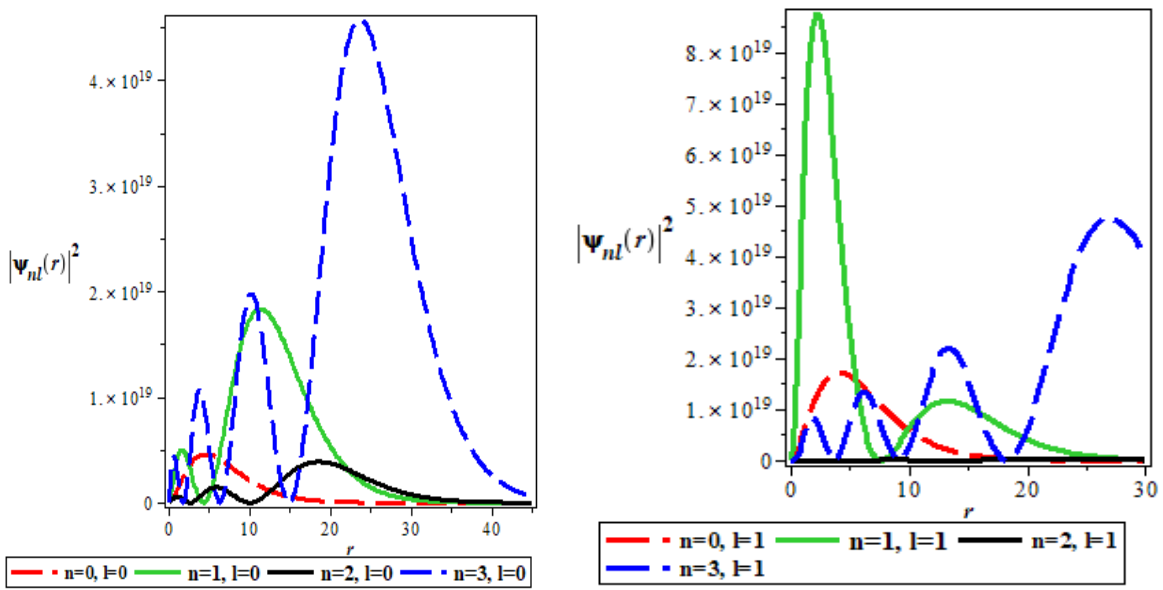

Figure 4: Variation of the probability density against the internuclear separation at various quantum state for $l=0$ and $l=1$ 333respectively for Yukawa potential.

## 4. Expectation Values Using Hellmann-Feynman Theorem

In this section, some expectation values are obtain using Hellmann-Feynman Theorem (HFT). According to Hellmann-Feynman Theorem, the Hamiltonian H for a particular quantum mechanical system is express as a function of some parameters $q$. Let $E(q)$ and $\Psi(q)$ be the eigenvalues and eigen function of the Hamiltonian. Then

$$\frac{\partial E_{nl}}{\partial q} = \left\langle \Psi(q) \left| \frac{\partial H(q)}{\partial q} \right| \Psi(q) \right\rangle. \tag{24}$$

For the purpose of clarity, the Hamiltonian for the propose potential using HFT is

$$H = -\frac{\hbar^2}{2\mu}\frac{d}{dr^2} + \frac{\hbar^2 l(l+1)}{2\mu r^2} - \frac{v_1}{r} + \left(\frac{B}{r} - \frac{v_2 \cosh\alpha}{r^2}\right)e^{-\alpha r}. \tag{25}$$

### 4.1 Expectation value for $<r^{-2}>_{nl}$

To obtain the expectation value for $<r^{-2}>_{nl}$, we set $q = l,$ to have

$$\langle r^{-2} \rangle_{n,l} = \alpha^2 - \frac{\alpha^2 (2l+1)}{2}\left[\frac{1}{2}(2l+1)\left(\sqrt{\left(l+\frac{1}{2}\right)^2 - \frac{2v_2\mu\cosh\alpha}{\hbar^2}}\right)^{-\frac{1}{2}} + Q_6\right]Q_7. \tag{26}$$

where

$$Q_6 = \frac{(2l+1)\left(n+\frac{1}{2}+\sqrt{\left(l+\frac{1}{2}\right)^2 - \frac{2v_2\mu\cosh\alpha}{\hbar^2}}\right) - \left(\frac{2\mu B}{\hbar^2\alpha} - \frac{2\mu v_1}{\hbar^2\alpha} + l(l+1)\right)\frac{1}{2}(2l+1)\left(\sqrt{\left(l+\frac{1}{2}\right)^2 - \frac{2v_2\mu\cosh\alpha}{\hbar^2}}\right)^{-\frac{1}{2}}}{\left(n+\frac{1}{2}+\sqrt{\left(l+\frac{1}{2}\right)^2 - \frac{2v_2\mu\cosh\alpha}{\hbar^2}}\right)^2}$$

$$Q_7 = \left\{\left(n+\frac{1}{2}+\sqrt{\left(l+\frac{1}{2}\right)^2 - \frac{2v_2\mu\cosh\alpha}{\hbar^2}}\right) + \frac{\left(\frac{2\mu B}{\hbar^2\alpha} - \frac{2\mu v_1}{\hbar^2\alpha} + l(l+1)\right)}{\left(n+\frac{1}{2}+\sqrt{\left(l+\frac{1}{2}\right)^2 - \frac{2v_2\mu\cosh\alpha}{\hbar^2}}\right)}\right\}$$

(27)

### 4.2 Expectation value for $<r^{-1}>_{nl}$

Setting $q = B$, then, the expectation value for $<r^{-1}>_{nl}$, becomes

$$<r^{-1}> = -\frac{\hbar^2\alpha^2 Q_7 \times e^{\alpha r}}{4\mu}\left[\frac{\left(\frac{2\mu}{\hbar^2\alpha^2}\right)\left(n+\frac{1}{2}+\sqrt{\left(l+\frac{1}{2}\right)^2-\frac{2v_2\mu\cosh\alpha}{\hbar^2}}\right)}{\left(n+\frac{1}{2}+\sqrt{\left(l+\frac{1}{2}\right)^2-\frac{2v_2\mu\cosh\alpha}{\hbar^2}}\right)^2}\right]. \qquad (28)$$

### 4.3 Expectation value for $<T>_{nl}$

Setting $q = \mu$, then

$$<T>_{nl} = \frac{\hbar^2\alpha^2 l(l+1)}{2\mu} - \frac{\hbar^2\alpha^2 Q_7^2}{8\mu}$$

$$+\frac{\hbar^2\alpha^2 Q_7}{4}\left[-\frac{v_2\cosh\alpha}{\hbar^2\sqrt{\left(l+\frac{1}{2}\right)^2-\frac{2v_2\mu\cosh\alpha}{\hbar^2}}}+\frac{\left(\frac{2}{\hbar^2\alpha}(B-v_1)\right)\left(\sqrt{\left(l+\frac{1}{2}\right)^2-\frac{2v_2\mu\cosh\alpha}{\hbar^2}}\right)+\left(\frac{v_2\cosh\alpha}{\hbar^2}\right)\left(\frac{2\mu B}{\hbar^2\alpha}-\frac{2\mu v_1}{\hbar^2\alpha}+l(l+1)\right)}{\left(n+\frac{1}{2}+\sqrt{\left(l+\frac{1}{2}\right)^2-\frac{2v_2\mu\cosh\alpha}{\hbar^2}}\right)^2}\right]. \qquad (29)$$

### 4.4 Expectation value for $<p^2>_{nl}$

The relationship between $T$ and $p^2$ is given as $T = \dfrac{p^2}{2\mu}$, therefore

$$<p^2>_{nl} = \hbar^2\alpha^2 l(l+1) - \frac{\hbar^2\alpha^2 Q_7^2}{4}$$

$$+ \frac{\hbar^2\alpha^2\mu Q_7}{2}\left[ -\frac{v_2\cosh\alpha}{\hbar^2\sqrt{\left(l+\frac{1}{2}\right)^2 - \frac{2v_2\mu\cosh\alpha}{\hbar^2}}} + \frac{\left(\dfrac{2}{\hbar^2\alpha}(B-v_1)\right)\left(\sqrt{\left(l+\frac{1}{2}\right)^2 - \dfrac{2v_2\mu\cosh\alpha}{\hbar^2}}\right) + \left(\dfrac{v_2\cosh\alpha}{\hbar^2}\right)\left(\dfrac{2\mu B}{\hbar^2\alpha} - \dfrac{2\mu v_1}{\hbar^2\alpha} + l(l+1)\right)}{\left(n+\frac{1}{2} + \sqrt{\left(l+\frac{1}{2}\right)^2 - \dfrac{2v_2\mu\cosh\alpha}{\hbar^2}}\right)^2} \right]. \qquad (30)$$

### 5 Thermodynamic Properties

In this section, we present the thermodynamic properties for the potential model. The thermodynamic properties of quantum systems can be obtained from the exact partition function given by

$$Z(\beta) = \sum_{n=0}^{\lambda} e^{-\beta E_n} \qquad (31)$$

where, $\lambda$ is an upper bound of the vibrational quantum number obtained from the numerical solution of $\dfrac{dE_n}{dn} = 0$, given as $\lambda = -\delta + \sqrt{Q_3}$, $\beta = \dfrac{1}{kT}$ where $k$ and T are Boltzmann constant and absolute temperature respectively. In the classical limit, the summation in equation (31) can be replaced with an integral:

$$Z(\beta) = \int_0^{\lambda} e^{-\beta E_n} dn. \qquad (32)$$

In order to obtain the partition function, the energy equation (14) can be presented in a close and compact form as

$$E_{nl} = \frac{\hbar^2\alpha^2 l(l+1)}{2\mu} - v_1\alpha - \frac{\hbar^2\alpha^2}{8\mu}\left\{\left(n+\frac{1}{2}+\sqrt{\left(l+\frac{1}{2}\right)^2 - \frac{2v_2\mu\cosh\alpha}{\hbar^2}}\right) + \frac{\left(\frac{2\mu B}{\hbar^2\alpha} - \frac{2\mu v_1}{\hbar^2\alpha} + l(l+1)\right)}{\left(n+\frac{1}{2}+\sqrt{\left(l+\frac{1}{2}\right)^2 - \frac{2v_2\mu\cosh\alpha}{\hbar^2}}\right)}\right\}^2 . \quad (33)$$

Equation (33) can further be simplified to

$$E_{nl} = Q_1 - Q_2\left\{(n+\delta) + \frac{Q_3}{(n+\delta)}\right\}^2, \quad (34)$$

where

$$Q_1 = \frac{\hbar^2\alpha^2 l(l+1)}{2\mu} - v_1\alpha, \quad Q_2 = \frac{\hbar^2\alpha^2}{8\mu}, \quad Q_3 = \left(\frac{2\mu B}{\hbar^2\alpha} - \frac{2\mu v_1}{\hbar^2\alpha} + l(l+1)\right), \quad \delta = \frac{1}{2} + \sqrt{\left(l+\frac{1}{2}\right)^2 - \frac{2v_2\mu\cosh\alpha}{\hbar^2}}$$

(35)

and equation (34) can be represented in the form

$$E_{nl} = Q_1 - Q_2\left[\rho + \frac{Q_3}{\rho}\right]^2 \Rightarrow -\left[Q_2\rho^2 + \frac{Q_2 Q_3^2}{\rho^2}\right] - (2Q_2 Q_3 - Q_1) \quad (36)$$

where,

$$\rho = n + \delta . \quad (37)$$

(i) Partition Function

Substituting equation (36) into equation (32) taking note of changes in the integration boundaries using equation (37) gives the partition function as

$$Z(\beta) = e^{\beta(2Q_2 Q_3 - Q_1)} \int_\delta^{\lambda+\delta} e^{\beta\left(Q_2\rho^2 + \frac{Q_2 Q_3^2}{\rho^2}\right)} d\rho . \quad (38)$$

Using Mathematica 10.0 version, the partition function of equation (38) is given as

$$Z(\beta) = \frac{e^{-\beta Q_1 + 2\beta Q_2 Q_3 - 2\aleph\sqrt{-\beta Q_2}}\sqrt{\pi}\left\{2\mathrm{erf}\left(\aleph_0 - \frac{\aleph}{\delta}\right)\right\}}{4\sqrt{-\beta Q_2}} . \quad (39)$$

(ii) Vibrational mean energy is given as

$$U(\beta) = -\frac{\partial \ln Z(\beta)}{\partial \beta} =$$

$$\left\{ \frac{2Q_2 e^{2\aleph\sqrt{-\beta Q_2}} \left[ -e^{\left(\frac{\beta Q_2(\delta^4+Q_3^2)}{\delta^2}\right)} \delta + e^{\left(\frac{\beta Q_2((\delta+\lambda)^4+Q_3^2)}{(\delta+\lambda)^2}\right)} \lambda + 2Q_2 e^{2\aleph\sqrt{-\beta Q_2}} \sqrt{\pi} erf \aleph \left( \aleph_2 - \aleph_0 - \frac{\aleph}{\delta} \right) \right]}{\sqrt{\pi} \left\{ erf \left( \aleph_0 - \frac{\aleph}{\delta} \right) - e^{4\aleph\sqrt{-\beta Q_2}} erf \left( \aleph_0 + \frac{\aleph}{\delta} \right) + erf \left( \aleph_1 - \frac{\aleph}{(\lambda+\delta)} \right) + e^{4\aleph\sqrt{-\beta Q_2}} erf \left( \aleph_1 + \frac{\aleph}{(\lambda+\delta)} \right) \right\}} \right\}. \quad (40)$$

(iii) Vibrational entropy is given as

$$S(\beta) = k \ln Z(\beta) - k\beta \frac{\partial \ln Z(\beta)}{\partial \beta}$$

$$= k \ln \frac{1}{4\sqrt{-\beta Q_2}} e^{-\beta Q_1 + 2\beta Q_2 Q_3 - 2\aleph\sqrt{-\beta Q_2}} \sqrt{\pi} \left\{ erf\left(\aleph_0 - \frac{\aleph}{\delta}\right) + e^{4\aleph\sqrt{-\beta Q_2}} erf\left(\aleph_1 + \frac{\aleph}{(\lambda+\delta)} - \aleph_0 - \frac{\aleph}{\delta}\right) + erf\left(\aleph_1 - \frac{\aleph}{(\lambda+\delta)}\right) \right\}$$

$$-k\beta \left\{ \frac{2Q_2 e^{2\sqrt{-\beta Q_2}} \sqrt{-\beta Q_2 Q_3^2} \left[ -e^{\left(\frac{\beta Q_2(\delta^4+Q_3^2)}{\delta^2}\right)} \delta + e^{\left(\frac{\beta Q_2((\delta+\lambda)^4+Q_3^2)}{(\delta+\lambda)^2}\right)} \lambda + 2Q_2 e^{2\aleph\sqrt{-\beta Q_2}} \sqrt{\pi} erf \aleph \left( \aleph_2 - \aleph_0 - \frac{\aleph}{\delta} \right) \right]}{\sqrt{\pi} \left\{ 2 erf \left( \aleph_0 - \frac{\aleph}{\delta} \right) \right\}} \right\}. \quad (41)$$

(iv) Vibrational Free energy is given as

$$F(\beta) = -\frac{1}{\beta} \ln Z(\beta)$$

$$= -\frac{1}{\beta} \ln \left( \frac{e^{-\beta Q_1 + 2\beta Q_2 Q_3 - 2\aleph\sqrt{-\beta Q_2}} \sqrt{\pi}}{4\sqrt{-\beta Q_2}} \left\{ erf \left( \aleph_1 - \frac{\aleph}{(\lambda+\delta)} + \aleph_0 - \frac{\aleph}{\delta} \right) + e^{4\aleph\sqrt{-\beta Q_2}} erf \left( \aleph_1 + \frac{\aleph}{(\lambda+\delta)} - \aleph_0 - \frac{\aleph}{\delta} \right) \right\} \right). \quad (42)$$

(v) Vibrational specific heat capacity is given as

$$C(\beta) = k\beta^2 \left( \frac{\partial^2 \ln Z(\beta)}{\partial \beta^2} \right) = k\sqrt{-\beta Q_2} e^{2\aleph\sqrt{-\beta Q_2}}$$

$$\times \left\{ \begin{array}{l} \left[ -erf(\Lambda_1) - e^{4\aleph\sqrt{-\beta Q_2}} erf(\Lambda_2) + erf(\Lambda_3) + e^{4\aleph\sqrt{-\beta Q_2}} erf(\Lambda_4) \right] \aleph_4 + \dfrac{\aleph_4^2}{4\sqrt{\pi}\sqrt{-\beta Q_2}} - \\[6pt] 4\aleph \left[ erf(\Lambda_1 - \Lambda_3) + e^{4\aleph\sqrt{-\beta Q_2}} erf(\Lambda_2 - \Lambda_4) \right] \sqrt{-\beta Q_2} \aleph_5 - \dfrac{\aleph_8 \aleph_6}{\left( \delta(\lambda+\delta)\sqrt{-\beta Q_2} \right)^2} \\[6pt] \dfrac{+\Lambda_5 \left( -\delta^2 \sqrt{-\beta Q_2} + 2\aleph \right) - 2\beta Q_2 \aleph \Lambda_6 + \delta(\lambda+\delta) \aleph_7}{\sqrt{\pi} \left[ erf(\Lambda_1) + e^{4\aleph\sqrt{-\beta Q_2}} erf(\Lambda_1) - erf(\Lambda_3) - e^{4\aleph\sqrt{-\beta Q_2}} erf(\Lambda_4) \right]^2} \end{array} \right\}. \quad (43)$$

where

$$\left. \begin{array}{l} \Lambda_1 = \aleph_0 - \dfrac{\aleph}{\delta}, \Lambda_2 = \aleph_0 + \dfrac{\aleph}{\delta}, \Lambda_3 = \aleph_1 - \dfrac{\aleph}{(\lambda+\delta)}, \Lambda_4 = \aleph_1 + \dfrac{\aleph}{(\lambda+\delta)}, \Lambda_5 = e^{\left( \frac{\beta Q_2(\delta^4 + Q_3^2)}{\delta^2} \right)} \delta, \\[8pt] \Lambda_6 = e^{\left( \frac{\beta Q_2(\delta+\lambda)^4 + Q_3^2}{(\delta+\lambda)^2} \right)} \delta,, \aleph = \sqrt{-\beta Q_2 Q_3^2}, \aleph_0 = \sqrt{-\beta Q_2}\delta, \aleph_1 = \sqrt{-\beta Q_2}(\lambda+\delta), \\[8pt] \aleph_2 = \aleph_0 \delta + 2\delta\lambda\sqrt{-\beta Q_2} + \dfrac{\lambda^2\sqrt{-\beta Q_2} + \aleph}{\lambda+\delta}, \aleph_3 = 2\aleph e^{2\aleph\sqrt{-\beta Q_2}}\sqrt{\pi}, \aleph_4 = -\Lambda_5 + 2\Lambda_6 + \aleph_3 erf(\Lambda_8 - \Lambda_1), \\[8pt] \aleph_5 = \Lambda_5 - 2\Lambda_6 + \aleph_3 erf(\Lambda_2 - \Lambda_8), \aleph_6 = Q_2^2 Q_3^2 - e^{2\aleph\sqrt{-\beta Q_2}}\sqrt{\pi}\delta(\lambda+\delta) erf(\Lambda_8 - \Lambda_2) \aleph\sqrt{-\beta Q_2}, \\[8pt] \aleph_7 = erf(\Lambda_6)\left( \delta^2\sqrt{-\beta Q_2} + 2\delta\lambda\sqrt{-\beta Q_2} + \lambda^2\sqrt{-\beta Q_2} - 2\aleph \right), \\[8pt] \aleph_8 = -erf(\Lambda_3) + e^{4\aleph\sqrt{-\beta Q_2}} erf(\Lambda_2 - \Lambda_3) - 4\sqrt{\pi}\beta^2 e^{4\aleph\sqrt{-\beta Q_2}}\delta(\lambda+\delta) erf(\Lambda_8) \end{array} \right\}. \quad (44)$$

## 6. Superstatistics Formulation

Superstatistics is the superposition of two different statistics which is applicable to driven non equilibrium systems to statistical intensive parameter ($\beta$) fluctuation [71]. This intensive parameter which undergoes spatio-temporal fluctuations include: chemical potential and energy fluctuation which is basically describe in terms of effective Boltzmann factor [72]. According to Edet et.al [73], the effective Boltzmann factor is given as

$$B(E) = \int_0^\infty e^{-\beta' E} f(\beta', \beta) d\beta', \tag{45}$$

where

$f(\beta', \beta) = \delta(\beta - \beta')$ is the Dirac delta function. However, the generalized Boltzmann factor express in terms of deformation parameter $q$ is given as

$$B(E) = e^{-\beta E}\left(1 + \frac{q}{2}\beta^2 E^2\right). \tag{46}$$

The partition function for superstatistics formalism is then given as

$$Z_s = \int_0^\infty B(E) dn. \tag{47}$$

Substituting equation (34) into equation (46) gives the generalized Boltzmann factor equation as

$$B(E) = \left[1 + \frac{q}{2}\beta^2\left(-\left(Q_2\rho^2 + \frac{Q_2 Q_3^2}{\rho^2}\right) - (2Q_2 Q_3 - Q_1)\right)^2\right] e^{-\beta\left[-\left(Q_2\rho^2 + \frac{Q_2 Q_3^2}{\rho^2}\right) - (2Q_2 Q_3 - Q_1)\right]}. \tag{48}$$

Using equation (47), the superstatistics partition function equation is given as

$$Z_s = e^{\beta(2Q_2 Q_3 - Q_1)} \int_0^\infty \left[1 + \frac{q}{2}\beta^2\left(-\left(Q_2\rho^2 + \frac{Q_2 Q_3^2}{\rho^2}\right) - (2Q_2 Q_3 - Q_1)\right)^2\right] e^{\beta\left(Q_2\rho^2 + \frac{Q_2 Q_3^2}{\rho^2}\right)} d\rho. \tag{49}$$

Using Mathematica 10.0 version, the partition obtain from equation (47) is

$$Z_s = \frac{1}{16\sqrt{-\beta Q_2 Q_3^2}\,\beta Q_2}\left\{e^{-\beta Q_1 + 2\beta Q_2 Q_3 - 2\sqrt{-\beta Q_2}\sqrt{-\beta Q_2 Q_3^2}}\sqrt{\pi}\begin{bmatrix}-32q\beta^3 Q_2^3 Q_3^3 \\ +\left(8 + 3q + 4q\beta Q_1 + 4q\beta^2 Q_1^2\right)\sqrt{-\beta Q_2}\sqrt{-\beta Q_2 Q_3^2} \\ +8q(1+2\beta Q_1)(-\beta Q_2)^{\frac{3}{2}}Q_3\sqrt{-\beta Q_2 Q_3^2} \\ +8q^2\beta^2 Q_2^2 Q_3^2 \times \left(1 + 2\beta Q_1 + 4\sqrt{-\beta Q_2}\sqrt{-\beta Q_2 Q_3^2}\right)\end{bmatrix}\right\}. \tag{50}$$

We use the same procedure of thermodynamic section 5 to obtain superstatistics vibrational mean energy ($U_s$), vibrational Specific heat capacity ($C_s$), Vibrational entropy ($S_s$) and vibrational free energy ($F_s$) from the partition equation (41). However, this solution is not included in the article because of the lengthy and bulky analytical equations.

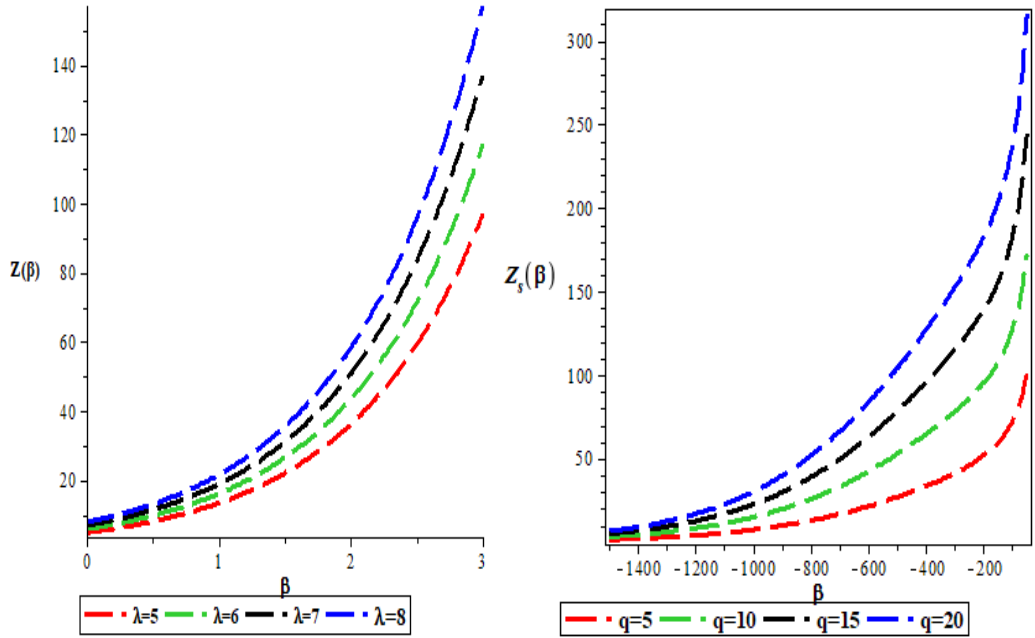

Figure 5: Variation of partition function with $\beta$ and q for thermodynamic properties and superstatistics respectively.

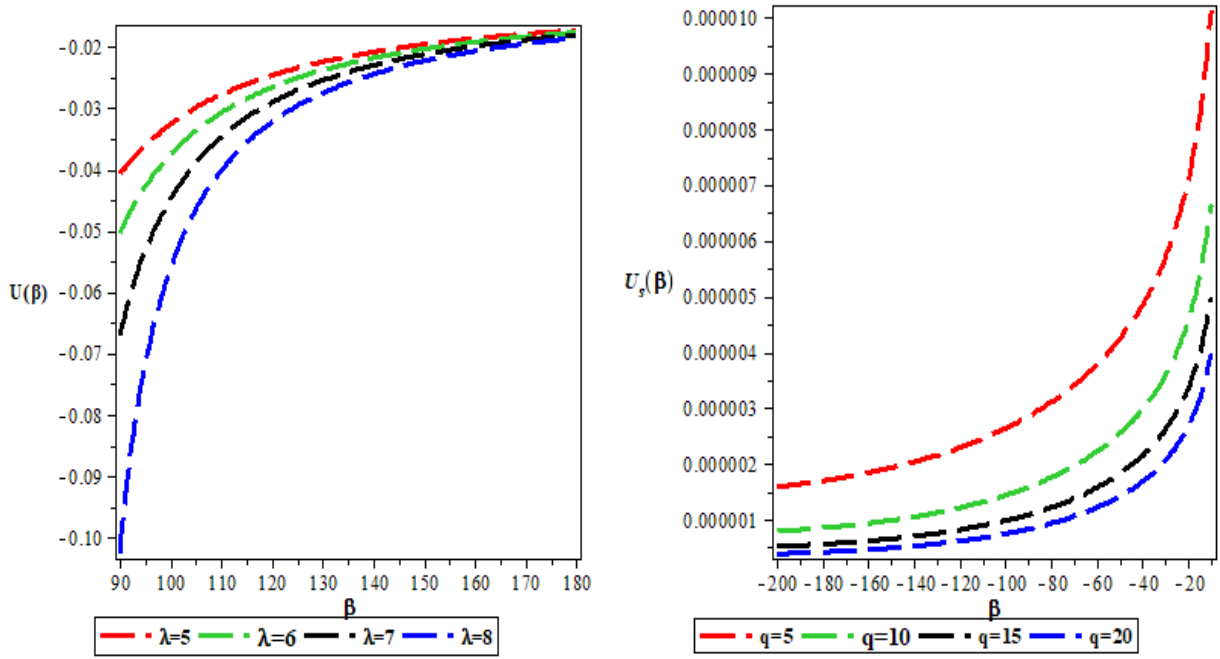

Figure 6: Variation of Vibrational mean energy with $\beta$ and q for thermodynamic properties and superstatistics respectively.

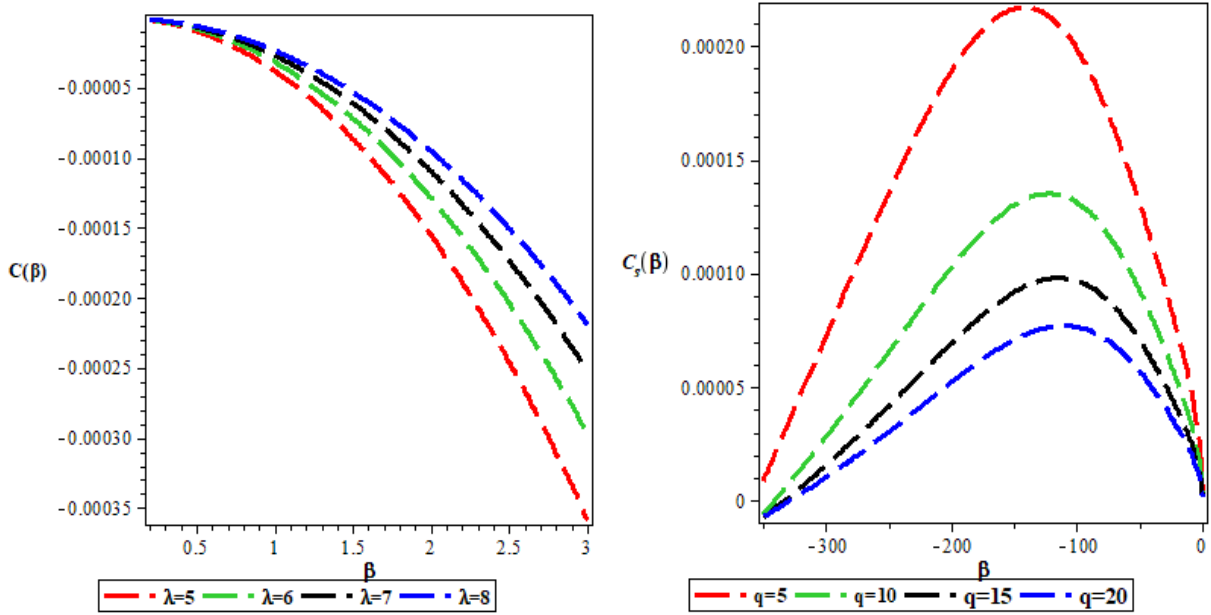

Figure 7: Variation of Vibrational specific heat capacity with $\beta$ and q for thermodynamic properties and superstatistics respectively.

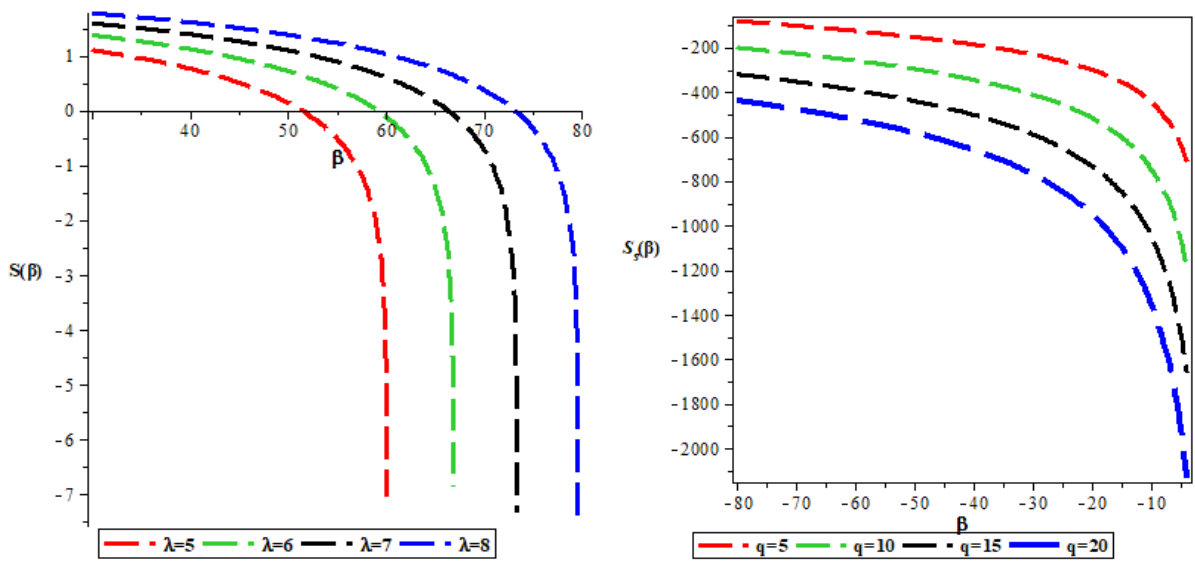

Figure 8: Variation of Vibrational entropy with $\beta$ and q for thermodynamic properties and superstatistics respectively.

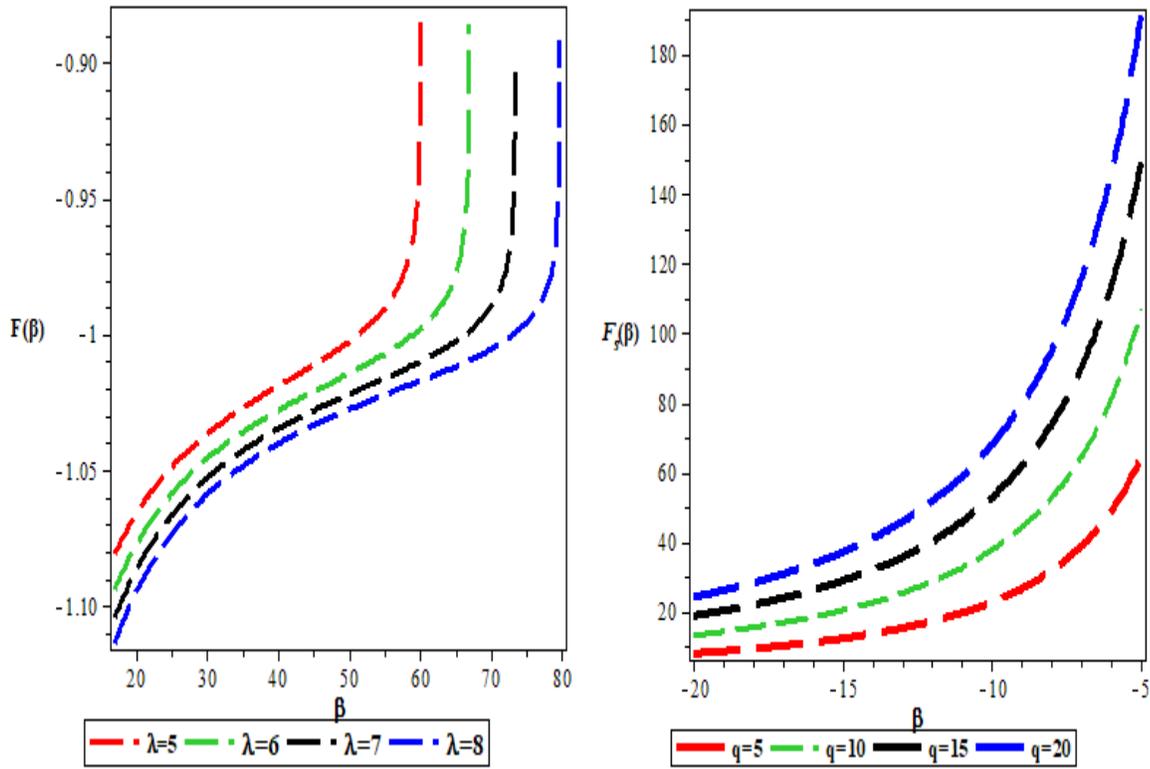

Figure 11 : Variation of Vibrational free energy with $\beta$ and q for thermodynamic properties and superstatistics respectively.

## 7 Numerical Results

Using Matlab 10.0 version, the numerical bound state solutions for the proposed potential was calculated using (14) for different quantum state. Also, using equations (18), (20), (21) and (22), the expectation values for $<r^{-2}>_{nl}$, $<r^{-1}>_{nl}$, $<T>_{nl}$ and $<p^2>_{nl}$ respectively were calculated as shown in tables 1 to 5.

Table 1. Numerical bound state solutions. $B = 0.2, v_1 = 0.1V$, $v_2 = 0.2V$, $\hbar = \mu = 1$

| n | l | $E_{nl}, \alpha = 0.01$ (eV) | $E_{nl}, \alpha = 0.02$ (eV) | $E_{nl}, \alpha = 0.03$ (eV) | $E_{nl}, \alpha = 0.04$ (eV) |
|---|---|---|---|---|---|
| 0 | 0 | -0.03061781 | -0.033096257 | -0.035560349 | -0.038010115 |
| 1 | 0 | -0.01893207 | -0.021509527 | -0.024138612 | -0.026819316 |
|   | 1 | -0.00842605 | -0.011467973 | -0.014626685 | -0.017902168 |
| 2 | 0 | -0.00927784 | -0.012006400 | -0.014887315 | -0.017920574 |
|   | 1 | -0.00586094 | -0.008981149 | -0.012380582 | -0.016059205 |

| n | l | | | | |
|---|---|---|---|---|---|
| | 2 | -0.00490303 | -0.008288097 | -0.012015183 | -0.016084269 |
| 3 | 0 | -0.00619246 | -0.009146159 | -0.012402307 | -0.015960892 |
| | 1 | -0.00477947 | -0.008148623 | -0.001201250 | -0.016371067 |
| | 2 | -0.00434223 | -0.008001341 | -0.012261886 | -0.017123836 |
| | 3 | -0.00414695 | -0.008039833 | -0.012612300 | -0.017864352 |
| 4 | 0 | -0.04924686 | -0.008178413 | -0.011934614 | -0.016193276 |
| | 1 | -0.00427845 | -0.008007979 | -0.012498900 | -0.017751159 |
| | 2 | -0.00408585 | -0.008147119 | -0.013123860 | -0.019016034 |
| | 3 | -0.00401576 | -0.008352259 | -0.013722838 | -0.020127466 |
| | 4 | -0.00400000 | -0.008568537 | -0.014266689 | -0.021094435 |
| 5 | 0 | -0.00434190 | -0.007970639 | -0.012351858 | -0.017485540 |
| | 1 | -0.00405927 | -0.008241005 | -0.013501346 | -0.019840219 |
| | 2 | -0.00400216 | -0.008568666 | -0.014417121 | -0.021547480 |
| | 3 | -0.00400991 | -0.008901268 | -0.015236763 | -0.023016365 |
| | 4 | -0.00404605 | -0.009216668 | -0.015965876 | -0.024293643 |
| | 5 | -0.00409483 | -0.009507448 | -0.016612192 | -0.025409045 |

Table 2. Expectation values for $<r^{-2}>_{nl}$. $B = 0.2, v_1 = 0.1V$, $v_2 = 0.2V$, $\hbar = \mu = 1$

| n | l | $<r^{-2}>_{nl}$ $\alpha = 0.01, (\dot{A}^{-2})$ | $<r^{-2}>_{nl}$ $\alpha = 0.03, (\dot{A}^{-2})$ | $<r^{-2}>_{nl}$ $\alpha = 0.03, (\dot{A}^{-2})$ | $<r^{-2}>_{nl}$ $\alpha = 0.04, (\dot{A}^{-2})$ |
|---|---|---|---|---|---|
| 0 | 0 | -0.180845600 | -0.19312890 | -0.20559450 | -0.21823660 |
| 1 | 0 | -0.007804320 | -0.01146716 | -0.01567526 | -0.02042717 |
| | 1 | 0.031534220 | 0.03193252 | 0.03112350 | 0.02910763 |
| 2 | 0 | -0.003148090 | -0.00654506 | -0.01084855 | -0.01605580 |
| | 1 | 0.010561720 | 0.00855289 | 0.00383228 | -0.00359942 |
| | 2 | 0.057017470 | 0.04310601 | 0.01225219 | -0.03554203 |
| 3 | 0 | -0.002633730 | -0.00636279 | -0.01136044 | -0.01762263 |
| | 1 | 0.004432820 | -0.00033896 | -0.00987264 | -0.02416721 |
| | 2 | 0.021856240 | -0.01200783 | -0.07495907 | -0.16699480 |
| | 3 | 0.051656280 | -0.03959544 | -0.20947110 | -0.45796650 |
| 4 | 0 | -0.002613291 | -0.00679560 | -0.01260871 | -0.02004732 |
| | 1 | 0.001361432 | -0.00701901 | -0.02276403 | -0.04587216 |
| | 2 | 0.003256901 | -0.05268207 | -0.01526563 | -0.02966622 |
| | 3 | 0.002156285 | -0.15152770 | -0.42284930 | -0.81180300 |
| | 4 | 0.000077152 | -0.30591300 | -0.84750470 | -1.62484500 |
| 5 | 0 | -0.002708980 | -0.00738565 | -0.01405538 | -0.02271165 |
| | 1 | -0.000934536 | -0.01380659 | -0.03719615 | -0.07110123 |
| | 2 | -0.010539190 | -0.09245381 | -0.23627330 | -0.44199310 |
| | 3 | -0.034490940 | -0.25675980 | -0.64214770 | -1.19064800 |
| | 4 | -0.074756000 | -0.52092780 | -1.29216600 | -2.38846200 |
| | 5 | -0.129402700 | -0.88828570 | -2.20300600 | -4.07355300 |

Table 3. Expectation values for $<r^{-1}>_{nl}$. $B = 0.2, v_1 = 0.1V$, $v_2 = 0.2V$, $\hbar = \mu = 1$

| n | l | $<r^{-1}>_{nl}$ $\alpha = 0.01, (\dot{A}^{-1})$ | $<r^{-1}>_{nl}$ $\alpha = 0.02, (\dot{A}^{-1})$ | $<r^{-1}>_{nl}$ $\alpha = 0.03, (\dot{A}^{-1})$ | $<r^{-1}>_{nl}$ $\alpha = 0.04, (\dot{A}^{-1})$ |
|---|---|---|---|---|---|
| 0 | 0 | -0.19225140 | -0.19688090 | -0.20138910 | -0.20577640 |
| 1 | 0 | -0.01142578 | -0.01191251 | -0.01239767 | -0.01288129 |
|   | 1 | -0.04285259 | -0.04902030 | -0.05517621 | -0.06132020 |
| 2 | 0 | -0.04962766 | -0.05456679 | -0.05949515 | -0.06441276 |
|   | 1 | -0.02577863 | -0.03141289 | -0.03703604 | -0.04264812 |
|   | 2 | -0.01862148 | -0.02483049 | -0.03102713 | -0.03721144 |
| 3 | 0 | -0.02857951 | -0.03354061 | -0.03849154 | -0.04343230 |
|   | 1 | -0.01810582 | -0.02350037 | -0.02888420 | -0.03425735 |
|   | 2 | -0.01440590 | -0.02023623 | -0.02605494 | -0.03186203 |
|   | 3 | -0.01315972 | -0.01949603 | -0.02581969 | -0.03213072 |
| 4 | 0 | -0.01947010 | -0.02444052 | -0.02940090 | -0.03435125 |
|   | 1 | -0.01401303 | -0.01927972 | -0.02453591 | -0.02978164 |
|   | 2 | -0.01188206 | -0.01748570 | -0.02307814 | -0.02865941 |
|   | 3 | -0.01069620 | -0.01662480 | -0.02254153 | -0.02844645 |
|   | 4 | -0.00997767 | -0.01620214 | -0.02241417 | -0.02861378 |
| 5 | 0 | -0.01475625 | -0.01973146 | -0.02469666 | -0.02965188 |
|   | 1 | -0.01157554 | -0.01676609 | -0.02194628 | -0.02711614 |
|   | 2 | -0.01025245 | -0.01570970 | -0.02115605 | -0.02659152 |
|   | 3 | -0.00949240 | -0.01522176 | -0.02093966 | -0.02664612 |
|   | 4 | -0.00902694 | -0.01501492 | -0.02099092 | -0.02695497 |
|   | 5 | -0.19225140 | -0.19688090 | -0.02117606 | -0.02737791 |

Table 4. Expectation values for $<T>_{nl}$. $B = 0.2, v_1 = 0.1V$, $v_2 = 0.2V$, $\hbar = \mu = 1$

| n | l | $<T>_{nl}$ $\alpha = 0.01, (eV)$ | $<T>_{nl}$ $\alpha = 0.02, (eV)$ | $<T>_{nl}$ $\alpha = 0.03, (eV)$ | $<T>_{nl}$ $\alpha = 0.04, (eV)$ |
|---|---|---|---|---|---|
| 0 | 0 | -0.026806820 | -0.025402520 | -0.02391260 | -0.022337650 |
| 1 | 0 | -0.015431540 | -0.013981940 | -0.01205726 | -0.009657998 |
|   | 1 | -0.006571382 | -0.007487332 | -0.00824097 | -0.008824540 |
| 2 | 0 | -0.005631741 | -0.003353308 | 0.00013309 | 0.004825102 |
|   | 1 | -0.004014817 | -0.004921358 | -0.00572940 | -0.006428803 |
|   | 2 | -0.002395157 | -0.002365251 | -0.00173894 | -0.000484917 |
| 3 | 0 | -0.002066055 | 0.001684466 | 0.00770820 | 0.015999570 |
|   | 1 | -0.002916120 | -0.003925660 | -0.00491908 | -0.005881812 |
|   | 2 | -0.001738771 | -0.001662435 | -0.00102536 | 0.002026249 |
|   | 3 | -0.001048483 | 0.000019672 | 0.00235305 | 0.006033901 |
| 4 | 0 | -0.000071586 | 0.005705386 | 0.01515297 | 0.028261000 |
| | 1 | -0.002372722 | -0.003544271 | -0.00480446 | -0.006132816 |
| | 2 | -0.001362593 | -0.001292758 | -0.00069870 | 0.000451413 |
| | 3 | -0.000758604 | 0.000354150 | 0.02700911 | 0.006357665 |
| | 4 | -0.002157670 | 0.002154345 | 0.00672056 | 0.013654220 |
| 5 | 0 | 0.001455576 | 0.009784670 | 0.02351393 | 0.042627300 |
| | 1 | -0.002090302 | -0.003469073 | -0.00506461 | -0.006849102 |
| | 2 | -0.001134154 | -0.001102620 | -0.00058791 | 0.000445057 |
| | 3 | -0.000568497 | 0.000558786 | 0.00289301 | 0.006508060 |
| | 4 | -0.000048447 | 0.002360339 | 0.00693053 | 0.013820340 |
| | 5 | 0.000506994 | 0.004490706 | 0.01188615 | 0.023002670 |

Table 5. Expectation values for $< p^2 >_{nl}$ . . $B = 0.2, v_1 = 0.1V$, $v_2 = 0.2V$, $\hbar = \mu = 1$

| n | l | $< p^2 >_{nl}$ $\alpha = 0.01, \left(\frac{eV}{c}\right)^2$ | $< p^2 >_{nl}$ $\alpha = 0.02, \left(\frac{eV}{c}\right)^2$ | $< p^2 >_{nl}$ $\alpha = 0.03, \left(\frac{eV}{c}\right)^2$ | $< p^2 >_{nl}$ $\alpha = 0.04, \left(\frac{eV}{c}\right)^2$ |
|---|---|---|---|---|---|
| 0 | 0 | -0.053613630 | -0.050805050 | -0.04782520 | -0.04467531 |
| 1 | 0 | -0.030863080 | -0.027963880 | -0.02411452 | -0.01931600 |
| | 1 | -0.013142760 | -0.014974660 | -0.01648195 | -0.01764908 |
| 2 | 0 | -0.011263480 | -0.006706617 | 0.00026618 | 0.00965020 |
| | 1 | -0.008029633 | -0.000984271 | -0.01145882 | -0.01285761 |
| | 2 | -0.004790315 | -0.004730502 | -0.00347789 | -0.00096983 |
| 3 | 0 | -0.004132109 | 0.003368932 | 0.015416410 | 0.03199915 |
| | 1 | -0.005832247 | -0.007851326 | -0.009838166 | -0.01176362 |
| | 2 | -0.003477542 | -0.003324870 | -0.002050733 | 0.00040524 |
| | 3 | -0.002096967 | 0.000393442 | 0.004706108 | 0.01206780 |
| 4 | 0 | -0.000143173 | 0.011410770 | 0.030305940 | 0.05652200 |
| | 1 | -0.004745445 | -0.007088541 | -0.000960893 | -0.01226563 |
| | 2 | -0.002725186 | -0.002585515 | -0.001397411 | 0.00090282 |
| | 3 | -0.001517209 | 0.000708300 | 0.005401821 | 0.01271533 |
| | 4 | -0.000431534 | 0.004308690 | 0.013441140 | 0.02730844 |
| 5 | 0 | 0.002911151 | 0.019569340 | 0.047027870 | 0.08525461 |
| | 1 | -0.004180604 | -0.006938146 | -0.010129220 | -0.01369820 |
| | 2 | -0.002268308 | -0.002205244 | -0.001175838 | 0.00089011 |
| | 3 | -0.001136990 | 0.001117574 | 0.005786028 | 0.01301610 |
| | 4 | -0.000096894 | 0.004720678 | 0.013861070 | 0.02764067 |
| | 5 | 0.001013988 | 0.008981412 | 0.023772290 | 0.04600535 |

### 7.1 Special cases

(a) Hellmann Potential: Substituting $v_2 = 0$ into equation (1), then the potential reduces to Hellmann potential

$$v(r) = -\frac{v_1}{r} + \frac{Be^{-\alpha r}}{r}. \tag{51}$$

The required energy equation is

$$E_{nl} = \frac{\hbar^2 \alpha^2 l(l+1)}{2\mu} - v_1 \alpha - \frac{\hbar^2 \alpha^2}{8\mu}\left\{(n+l+1) + \frac{\left(\frac{2\mu}{\hbar^2 \alpha}(B-v_1) + l(l+1)\right)}{(n+l+1)}\right\}^2. \tag{52}$$

(b) Yukawa potential: If $v_1 = v_2 = 0$, then equation (1) reduces to Yukawa potential

$$v(r) = \frac{Be^{-\alpha r}}{r}. \tag{53}$$

The corresponding energy eigen equation is

$$E_{nl} = \frac{\hbar^2 \alpha^2 l(l+1)}{2\mu} - \frac{\hbar^2 \alpha^2}{8\mu}\left\{(n+l+1) + \frac{\left(\frac{2\mu B}{\hbar^2 \alpha} + l(l+1)\right)}{(n+l+1)}\right\}^2. \tag{54}$$

(c) Screened-Hyperbolic Inversely Quadratic Potential.

Substituting $B = v_1 = 0$, into equation (1), the potential reduces to screened-hyperbolic inversely quadratic potential

$$v(r) = \frac{v_2 e^{-\alpha r} \cosh \alpha}{r^2}. \tag{55}$$

The resulting energy eigen equation is

$$E_{nl} = \frac{\hbar^2 \alpha^2 l(l+1)}{2\mu} - \frac{\hbar^2 \alpha^2}{8\mu} \left\{ \left( n + \frac{1}{2} + \sqrt{\left(l + \frac{1}{2}\right)^2 - \frac{2v_2 \mu \cosh \alpha}{\hbar^2}} \right) + \frac{l(l+1)}{\left( n + \frac{1}{2} + \sqrt{\left(l + \frac{1}{2}\right)^2 - \frac{2v_2 \mu \cosh \alpha}{\hbar^2}} \right)} \right\}^2. \quad (56)$$

(d) Coulomb Potential: Substituting $\alpha = 0$, into equation (53), then, the potential reduces to Coulomb potential

$$v(r) = \frac{B}{r}. \quad (57)$$

By substituting $\alpha = 0$, into equation (54) gives the corresponding energy eigen equation for Coulomb's potential as

$$E_{nl} = -\frac{\hbar^2 \mu B^2}{2(n+l+1)^2}. \quad (58)$$

## 8 Discussion

Figure 1 is the graph of Pekeries approximation against the screening parameter $\alpha$. This graph shows that the approximation is suitable for the proposed potential. Variation of the probability density against the internuclear separation at various quantum state for $l=0$ and $l=1$ respectively are shown in figure 2. The variation of the probability are similar but for $l=0$, there is a more concentration of the electron density at the origin for all the quantum state studied. The concentration is higher for $l=0$. At every value of the internuclear distance, the probability density for $l=0$ is higher than the probability density for $l=1$. Figure 3: Variation of the probability density against the internuclear separation at various quantum state for $l=0$ and $l=1$ respectively for Hellmann potential is presented. A more concentration of the electron density is observed at the origin in both cases. It is also seen that the probability density obtained for $l=0$ are lower than the probability density obtained for $l=1$. In figure 4, we presented the variation of the probability density against the internuclear separation at various quantum state for $l=0$ and $l=1$ respectively for Yukawa potential. There is more concentration of the electron density at the origin for $l=0$, but this situation is not the same for $l=1$. However, the probability density at the second excited state for $l=1$, remains constant for all values of the internuclear separation while for $l=0$, the revise is the case.

In Figure 5, the variation partition function for non-superstatistics and superstatistics with the temperature parameter were observed. The partition function increases non-linearly with β. In both cases, the partition function diverged as β increases, but later converges for the superstatistics. The partition function converges as β becomes positive in the superstatistics. In Figure 6, we presented the variation of vibrational mean energy against $\beta$ for thermodynamic properties and superstatistics respectively. For the non-superstatistics, the mean energy increases as β goes up for all values of λ. However, at higher values of β, the mean energy for various λ tends to converge. The superstatistics mean energy rises as the temperature of the system decreases. However, at a certain absolute temperature, the superstatistics mean energy increases vertically for various values of the deformed parameter. The variation are opposite. In Figure 7, the variation of the heat capacity against β for four values of λ and q are shown. In each case, the heat capacity decreases monotonically with an increasing β for the non-superstatistics. At zero value of β, the heat capacity for various λ converged and diverge as β increases gradually. For the superstatistics, the heat capacity for various deformed parameter rises while the temperature cools down. The specific heat capacity has a turning point when β equals -150, the specific heat capacity for the superstatistics has a turning point and then converged at absolute zero. In figure 8, the vibrational entropy decreases and diverged while the temperature of the system decreases (β increases) in the non-superstatistics. This decrease is sharper for negative values of the entropy when β is almost constant. The superstatistics entropy varies inversely with β (directly with temperature). This means that when the temperature of the system is raised, the disorderliness of the system also increases for every value of the deformed parameter. The entropy for the superstatistics converged as the temperature parameter tends to zero. In Figure 10, the variation of free energy against the β is seen to be two different steps in the case of a non-superstatistics. Between 0 and 60 values of β, the free energy increases steadily for the various values of λ but beyond this range, the free energy increases sharply at constant β. For the superstatistics, the free energy increases monotonically as the temperature of the system reduces gradually. The free energy is always higher when the deformed parameter is increased.

Table 1 is the numerical bound state solutions for CPSEHP. The numerical bound state solutions increases with an increase in quantum state, but decreases with an increase in the screening parameter. Tables 2, 3, 4, and 5 are expectation values for $<r^{-2}>_{nl}$, $<r^{-1}>_{nl}$, $<T>_{nl}$ and $<p^2>_{nl}$ respectively. Here, the numerical values in all cases decreases with an increase in the screening parameter.

## 9 Conclusion

In this work, we apply the parametric Nikiforov-Uvarov method to obtain the bound state solutions of Coulomb plus screened-exponential hyperbolic potential. The resulting energy eigen equation were presented in a close and compact form. The research work was extended to study thermal properties, superstatistics and various expectation values. The propose potential also reduce to Hellmann potential, Yukawa potential, Screened Hyperbolic potential and Coulomb potential as special cases. The normalized wave function for the mother potential and that of the Hellmann potential are similar but the normalized wave function of the Yukawa potential seams different. The trend of the thermodynamic and superstatistics curves are in agreement to results of an existing literature. The results of the thermodynamic properties and superstatistics revealed that the effect of the temperature on the thermodynamic properties and the superstatistics are similar. Finally, this research work has practical applications in Physical and chemical sciences.

Data Availability: The data used for this work are generated using Matlab programme

**References**.